\documentclass[draft]{agujournal}
\draftfalse

\journalname{Geophysical Research Letters}

\usepackage{url}
\begin{document}

%% ------------------------------------------------------------------------ %%
%  Title
% 
% (A title should be specific, informative, and brief. Use
% abbreviations only if they are defined in the abstract. Titles that
% start with general keywords then specific terms are optimized in
% searches)
%
%% ------------------------------------------------------------------------ %%

% Example: \title{This is a test title}

\title{Fast response of deep ocean circulation to mid-latitude winds in the Atlantic}

%% ------------------------------------------------------------------------ %%
%
%  AUTHORS AND AFFILIATIONS
%
%% ------------------------------------------------------------------------ %%

% Authors are individuals who have significantly contributed to the
% research and preparation of the article. Group authors are allowed, if
% each author in the group is separately identified in an appendix.)

% List authors by first name or initial followed by last name and
% separated by commas. Use \affil{} to number affiliations, and
% \thanks{} for author notes.  
% Additional author notes should be indicated with \thanks{} (for
% example, for current addresses). 

% Example: \authors{A. B. Author\affil{1}\thanks{Current address, Antarctica}, B. C. Author\affil{2,3}, and D. E.
% Author\affil{3,4}\thanks{Also funded by Monsanto.}}

\authors{E. Frajka-Williams\affil{1}, F. Landerer\affil{2}, T. Lee\affil{2}}

\affiliation{1}{University of Southampton, National Oceanography Centre Southampton, Empress Dock, SO14 3ZH, United Kingdom}
\affiliation{2}{NASA Jet Propulsion Laboratory, Pasadena, CA, USA}
% \affiliation{3}{Third Affiliation}
% \affiliation{4}{Fourth Affiliation}

%% Corresponding Author:
% Corresponding author mailing address and e-mail address:

% (include name and email addresses of the corresponding author.  More
% than one corresponding author is allowed in this LaTeX file and for
% publication; but only one corresponding author is allowed in our
% editorial system.)  

% Example: \correspondingauthor{First and Last Name}{email@address.edu}

\correspondingauthor{E. Frajka-Williams}{e.frajka-williams@soton.ac.uk}

%% Keypoints, final entry on title page.

%  List up to three key points (at least one is required)
%  Key Points summarize the main points and conclusions of the article
%  Each must be 100 characters or less with no special characters or punctuation 

\begin{keypoints}
\item Covariance between GRACE and winds in the Atlantic identifies wind-driven changes of basinwide deep ocean circulation.
\item Atlantic MOC reversals in 2009/10 and 2010/11 resulted from the strongly negative NAO and mid-latitude wind stress curl.
\item Residual  interannual fluctuations in deep ocean transports are captured by GRACE satellite estimates of ocean bottom pressure.
\end{keypoints}

%% ------------------------------------------------------------------------ %%
%
%  ABSTRACT
%
% A good abstract will begin with a short description of the problem
% being addressed, briefly describe the new data or analyses, then
% briefly states the main conclusion(s) and how they are supported and
% uncertainties. 
%% ------------------------------------------------------------------------ %%

%% \begin{abstract} starts the second page 

\begin{abstract}
\textit{In situ} observations of transbasin deep ocean transports at $26^\circ$N show variability on monthly to decadal timescales (2004--2015).  Satellite-based estimates of ocean bottom pressure from the Gravity Recovery and Climate Experiment (GRACE) satellites were previously used to estimate interannual variability of deep ocean transports at $26^\circ$N.  Here, we use GRACE ocean bottom pressure, reanalysis winds and \textit{in situ} transport estimates at $26^\circ$N to diagnose the large-scale response of the deep ocean circulation to wind-forcing.  We find that deep ocean transports---including those associated with a reversal of the Atlantic meridional overturning circulation in 2009/10 and 2010/11---are part of a large-scale response to wind stress curl over the intergyre-gyre region.  Wind-forcing dominates deep ocean circulation variability on monthly timescales, but interannual fluctuations in the residual \textit{in situ} transports (after removing the wind-effect) are also captured by GRACE bottom pressure measurements.  On decadal timescales, uncertainty in regional trends in GRACE ocean bottom pressure preclude investigation of decadal-timescale transport trends.
\end{abstract}

%% ------------------------------------------------------------------------ %%
%
%  TEXT
%
%% ------------------------------------------------------------------------ %%@@\\
\section{Introduction}

Ocean circulation responds to forcing on a wide range of timescales. Century and longer duration simulations and paleoclimate records anticipate variations of the Atlantic meridional overturning circulation (AMOC) forcing or responding to climate changes \citep{Zhang-2008,LynchStieglitz-2017}.  Observations of monthly-to-interannual fluctuations in transbasin transports in the subtropical North Atlantic are largely goverend by wind-forcing \citep{Zhao-Johns-2014}.   While shorter timescale fluctuations may have less influence on climate timescales, they occur within the recent satellite observational period, enabling diagnosis of the basinscale response of ocean circulation to external forcing.

Since 2004, the AMOC has been measured at $26^\circ$N using a combination of moored and cable measurements by the RAPID Climate Change/Meridional Ocean Circulation and Heat flux Array (RAPID/MOCHA, hereafter RAPID) experiment \citep{McCarthy-etal-2015}.   These transport measurements show variability on monthly to interannual timescales \citep{Chidichimo-etal-2009,Kanzow-etal-2010,McCarthy-etal-2012,Smeed-etal-2014} including strong correlations between deep transports (3000--5000 m) and surface Ekman transport \citep{FrajkaWilliams-etal-2016}. Over the past two decades, the North Atlantic Oscillation (NAO) index has shown strongly anomalous values.  In the 2009/10 and 2010/11 winters, the NAO index was sharply negative.  The reorganisation of atmospheric winds during these periods (a southward shift of the position of the zero wind stress curl line) forced a reversal of surface meridional Ekman transport at $26^\circ$N and through it, a temporary reversal in the sign of the AMOC \citep{McCarthy-etal-2012} which repeated again in March 2013. All three events are characterised by a temporarily northward flowing North Atlantic Deep Water (NADW) layer \citep{FrajkaWilliams-etal-2016}, a watermass that is traditionally expected to flow southward in the deep western boundary current (DWBC) as the lower limb of the AMOC.

Concurrent with the RAPID observations, the GRACE (Gravity Recovery and Climate Experiment) satellites recorded spatial and temporal  variations in the Earth's distribution of mass.  Mass redistribution in the ocean drives circulation changes through geostrophy--whereby horizontal gradients in mass (or pressure) drive ocean transports normal to the gradient.  GRACE observations identified a large-scale gain and loss of mass in the intergyre-gyre region under the effect of negative or clockwise wind stress curl (WSC) \citep{Piecuch-Ponte-2014}.  The patterns of WSC are closely governed by sea level pressure anomalies and large-scale patterns are well-described by the NAO index.  In \cite{Piecuch-Ponte-2014}, they attributed 46\% of the nonseasonal ocean mass variations to a response to the WSC anomalies.
%\citep{Marshall-etal-2001} 

Changes in the strength of the southward transport of NADW are associated with a reduction of the overturning circulation.  The declining tendency of the $26^\circ$N AMOC is primarily contained in the reduction of the lower layer transports including the NADW \citep{Smeed-etal-2014}.  While transport variability in the deepest transport laters is derived primarily from a residual in the RAPID method \citep{McCarthy-etal-2012,FrajkaWilliams-etal-2016}, independent \textit{in situ} measurements of bottom pressure gradients confirm the RAPID estimates of deep transport variability on sub-annual  \citep{Kanzow-etal-2007,McCarthy-etal-2012}.  \textit{In situ} bottom pressure sensors are unable to measure decadal-scale changes due to intrinsic drift \citep{Watts-Kontoyiannis-1990}.  \cite{Landerer-etal-2015} used the GRACE estimates of ocean bottom pressure to independently determine the strength of the deep ocean transports from zonal gradients in bottom pressure, but due to uncertainties in separating long timescale GRACE bottom pressure signals from other gravity signals (e.g., glacial isostatic adjustment or GIA) did not evaluate trends in the deep ocean transports. As regional bottom pressure trends from GRACE are still uncertain, we will focus on the detrended GRACE values only. 

Here we use GRACE bottom pressure to diagnose the basinscale spatial fluctuations in the Atlantic on timescales less than a decade, and relate them to changes in the  deep ocean circulation.    While \citet{FrajkaWilliams-etal-2016} associated the deep transport variations at $26^\circ$N with a local reversal of the zonally-averaged wind stress, these satellite observations show instead that most of the transport variability in the lower NADW layer (3000--5000 m) at $26^\circ$N can be traced to a large-scale response of the ocean to anomalies in the WSC centered over the intergyre-gyre region, mediated by bottom topography.   %On longer timescales (10-year trends), the transport estimates from GRACE are subject to trends in ocean bottom pressure which are inconsistent with \textit{in situ} transport estimates (see \S2), and cannot be compared with \textit{in situ} bottom pressure records due to drift inherent in bottom pressure recorders.

%%% Suggested section heads:
% \section{Introduction}
% 
% The main text should start with an introduction. Except for short
% manuscripts (such as comments and replies), the text should be divided
% into sections, each with its own heading. 

% Headings should be sentence fragments and do not begin with a
% lowercase letter or number. Examples of good headings are:

% \section{Materials and Methods}
% Here is text on Materials and Methods.
%
% \subsection{A descriptive heading about methods}
% More about Methods.
% 
% \section{Data} (Or section title might be a descriptive heading about data)
% 
% \section{Results} (Or section title might be a descriptive heading about the
% results)
% 
% \section{Conclusions}

\section{Data and Methods}

We use monthly bottom pressure anomalies ($p_b$) grids over the period April 2002--June 2016 derived from GRACE time-variable gravity observations \citep{Tapley-etal-2004}. Specifically, we use the mascon solution from NASA's Jet Propulsion Laboratory (RL05M\_1.MSCNv02CRIv02; \citep{Watkins-etal-2015,Wiese-etal-2016}). The $p_b$ values are provided on a $1/2$ degree grid, with an effective spatial resolution of approximately 300 km.    Throughout the paper, we refer to values of $p_b$ in units of equivalent seawater thickness.  Maps of monthly 10-m wind fields defined on a regular 2$^\circ$ grid are obtained from the NCEP Reanalysis fields over the same period as the GRACE data (April 2002--June 2016). We use the reanalysis winds to compute wind stress with a variable drag coefficient updated for low wind speeds \citep{Large-Pond-1981,Trenberth-etal-1990}.  The monthly principal component-based index for the NAO was downloaded from \url{https://climatedataguide.ucar.edu/climate-data/hurrell-north-atlantic-oscillation-nao-index-pc-based}.
 
RAPID transport time series are used for the period April 2004--October 2015 \citep{Smeed-etal-2016}.   These are provided as 12-hourly transbasin (zonally-integrated) meridional transports which were estimated from submarine cable measurements of the Florida Current transport,  Ekman transport from reanalysis winds, and \textit{in situ} geostrophic transports from a mooring array across $26^\circ$N \citep{McCarthy-etal-2015}.  The deep ocean transports between the Bahamas and Canary Islands are separated into two layers: the upper North Atlantic Deep Water (NADW) transport from 1100--3000 m, and lower NADW transport from 3000--5000 m.

\begin{figure}[htb!]
\centering\includegraphics[width=\textwidth]{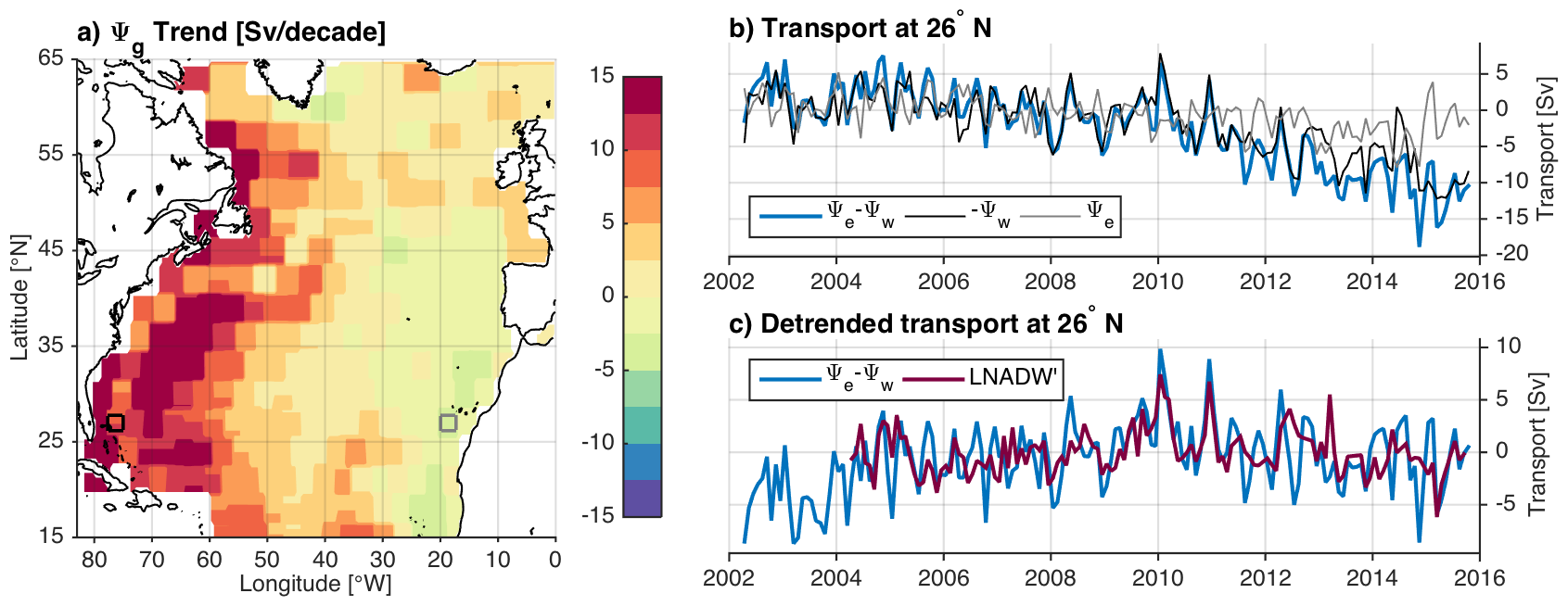}%ig1_trend_lay2_20180327.png}
\caption{(a) Trend in bottom transports estimated from GRACE bottom pressure over the period April 2004--October 2015.  (b) Transport anomalies at $26^\circ$N calculated from bottom pressure at the west (black, mascon 1271 location indicated by the black square in (a)), east (grey, mascon 1288 location indicated by the grey square in (a)) and the difference between them (blue).   (c) Detrended transport calculated from bottom pressure (blue) and from detrended lower NADW transport from the RAPID array (purple).}\label{fig:transports}
\end{figure}

Seasonal cycles were removed from all time series, calculated as the monthly climatology over the period April 2004--October 2015 (and, for spatial fields, at each pixel).  GRACE data were further processed by masking out the Hudson Bay, Gulf of Mexico and Caribbean, and filling gaps in time using linear interpolation. These gaps resulted from power-management on the satellites and are more common in the latter half of the record.   For clearer comparisons between GRACE and RAPID transports, gaps were created in the lower NADW transport and linearly interpolated.   To compare GRACE $p_b$, in units of centimeters liquid water equivalent, with transports measured at $26.5^\circ$N, bottom pressure fields are scaled to create a geostrophic streamfunction $\Psi_g$ as
\begin{equation}
\Psi_g = \frac{gp_b}{f}H\label{eq:streamfunction}
\end{equation}
where $g$ is the gravitational acceleration, $H$ the layer thickness, $p_b$ the bottom pressure in units of meters liquid water equivalent, and $f$ the Coriolis frequency.  Here, since GRACE bottom pressure is in units of height, the equation includes $gp_b[\mbox{m}]$ in place of  $p_b[\mbox{Pa}]/\rho_0$.  For comparison with the lower NADW transport from RAPID, we use a layer thickness $H$ of 2000 m, and  east minus west differences in pressure, where mascon 1271 (centered at $27^\circ$N, $76.42^\circ$W) and 1288 (centered at $27^\circ$N, $18.68^\circ$W) are used to represent the west and east side of the basin, respectively (Fig.~S1).

The GRACE bottom pressure has a long term background trend (Fig.~1a).  Due to a strong trend towards more negative bottom pressure anomalies in the east, a southward trend in transbasin ocean transports is implied where at $27^\circ$N, the trend in the east-west pressure gradient implies a transport trend  exceeding 10 Sv/decade.   Due to uncertainties in separating long timescale GRACE bottom pressure signals from other gravity signals (e.g., GIA) \citep{Landerer-etal-2015} did not evaluate trends in the deep ocean transports from GRACE. As regional OBP trends from GRACE are still uncertain, we focus here on the detrended values only where a linear trend over April 2004--2015 was removed at each pixel.

\section{Results}

\begin{figure}[htb!]
\centering\includegraphics[width=1\textwidth]{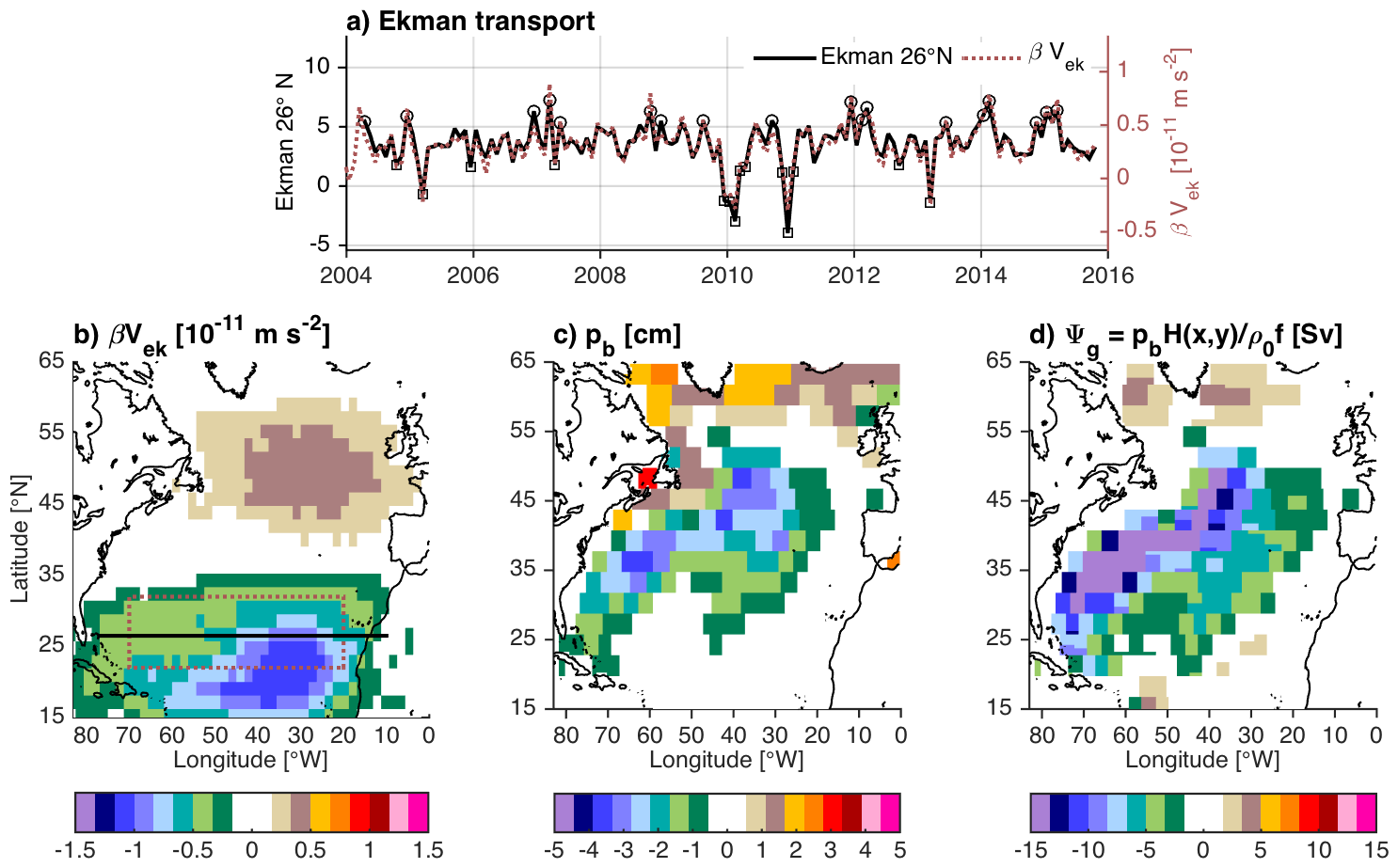}%{GRACE_like_Yeager13_new.png}
\caption{Similar to Fig. 13 in \citet{Yeager-2015}, but generated from  NCEP winds, RAPID transports and GRACE ocean bottom pressure.  (a) Time series of Ekman transport at $26.5^\circ$N (black) and local $\beta V_{ek}$ averaged over the red box in (b). Circles (squares) represent months included in the positive (negative) anomaly.  (b)--(d) show the composites for the mean of the negative months minus the positive months, where (b) shows $\beta V_{ek}$ and the latitude of the RAPID array (black), (c) the bottom pressure anomaly from GRACE, and (d) the streamfunction $\Psi_g$.}\label{fig:Yeager}
\end{figure}

Previous investigations identified a strong correlation between detrended and lowpass-filtered RAPID lower NADW transports and GRACE-derived bottom pressure gradients at $27^\circ$N in the Atlantic over the period April 2004--April 2014 \citep{Landerer-etal-2015}.  Here we show that this correlation holds on monthly timescales (Fig.~\ref{fig:transports}c).  At $26^\circ$N, there is a strong anti-correlation between lower NADW transports and meridional Ekman transport  \citep{FrajkaWilliams-etal-2016}.  \citet{Yeager-2015} uses annual $26^\circ$N Ekman transports to define positive and negative composites of circulation changes in a numerical model  (their Fig.~13).  They associate deep circulation anomalies with Ekman reversals, but more generally identify that local reversals at $26^\circ$N are associated with larger-scale changes in wind stress curl (WSC).  Here we repeat the composite analysis, using  Ekman transport at $26^\circ$N to identify months when Ekman transports are anomalous by greater than 1 standard deviation (Fig.~\ref{fig:Yeager}a).  Using these time periods, we calculate the difference between the mean of anomalies during negative months minus anomalies during positive months to identify basin-scale changes in wind-forcing, bottom pressure and ocean circulation (Fig.~\ref{fig:Yeager}b--d).  

As in \citet{Yeager-2015}, reversals in local $26^\circ$N Ekman transport coincide with a large-scale changes in curl.  Here the southward anomaly in the subtropics coincides with a weaker positive anomaly in the subpolar gyre, resulting in a divergence over the mid-latitudes (35--45$^\circ$N, Fig.~\ref{fig:Yeager}b).  This divergence results in a reduction in ocean bottom pressure (Fig.~\ref{fig:Yeager}c) but not confined to the mid-latitudes.  Rather, the anomaly extends to the south and west along the western side of the Atlantic basin (20--50$^\circ$N).  Scaling the bottom pressure anomalies through geostrophy, we find a barotropic streamfunction composite with an implied cyclonic circulation  (Fig.~\ref{fig:Yeager}d).  (Note that  the composites are derived here from monthly values, yielding larger magnitude anomalies than \citet{Yeager-2015}.)  At $26^\circ$N, these large-scale changes manifest as a northward anomaly in deep ocean transports, which is the signal captured in the RAPID array.

%For example, during the 2009/10 and 2010/11 winters, the lower NADW shows a northward transport anomaly (Fig.~\ref{fig:transports}c) coincident with the negative NAO during these same months.  The response was not a purely barotropic

\begin{figure}[hbt!]
\centering\includegraphics[width=.7\textwidth]{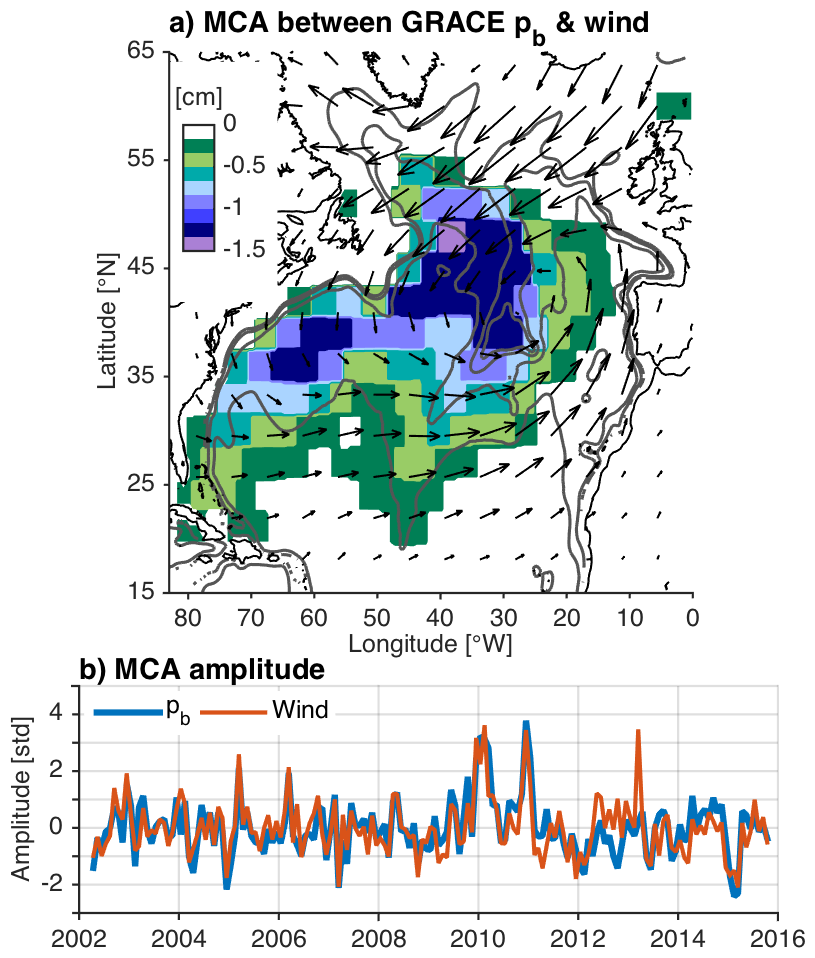}%{MCA_pattern_homocorr_20180327.png}
\caption{Maximum covariance analysis between the GRACE $p_b$ and zonal and meridional wind stress from NCEP.  The pattern of maximum covariance in GRACE $p_b$ is shaded, while the pattern for winds is indicated by the vectors.  Pixels are only colored where the correlation of the time series for GRACE in (b) is correlated with the original GRACE $p_b$ at that location with an $r\geq0.5$.  Vectors are  drawn where the correlation of the MCA for wind in (b) is correlated with either the $u$ or the $v$ time series at that pixel with $r\geq.3$.  (b) Time series of the pattern in (a) projected onto the GRACE $p_b$ data (blue) and  winds (red).  Note that the month of March 2013 (a peak in winds but not GRACE) occurred when the GRACE satellite had been in power-saving mode.  By the processing used here, the gap was linearly interpolated.}
\label{fig:MCA}
\end{figure}

While the composite analysis identifies coincident anomalies, it can be dominated by large amplitude anomalies with specific characteristic patterns.  Maximum covariance analysis (MCA) identifies patterns which covary in time, without setting a region of interest \textit{a priori}.  It has the potential to identify zero-lag relationships between wind-forcing and ocean response.  It has been used previously in the Atlantic by \citet{Piecuch-Ponte-2014} who identified nonseasonal fluctuations in wind stress curl and ocean mass changes.  We update it here with a higher-resolution GRACE product (the mascons version) and using a larger domain (15--65$^\circ$N, 83--0$^\circ$W) which includes the RAPID latitude (Fig.~\ref{fig:MCA}a). The MCA  identifies a center of wind action over $30^\circ$W, 40$^\circ$N with a strong resemblance to the North Atlantic Oscillation (NAO) pattern.  The MCA for bottom pressure shows that anomalies follow contours of planetary vorticity ($f/H$, where $f$ is the local Coriolis frequency and $H$ the local water depth, computed after smoothing bathymetry with a 300~km spatial filter) rather than contours of bathymetry (Fig.~S2).  The time series of variations, determined by projecting the spatial pattern onto the original space-time datasets, shows a high degree of correlation (by construction), but for smaller amplitude fluctuations as well as the big events in 2009/10 and 2010/11.

\begin{figure}[hbt!]
\centering\includegraphics[width=.8\textwidth]{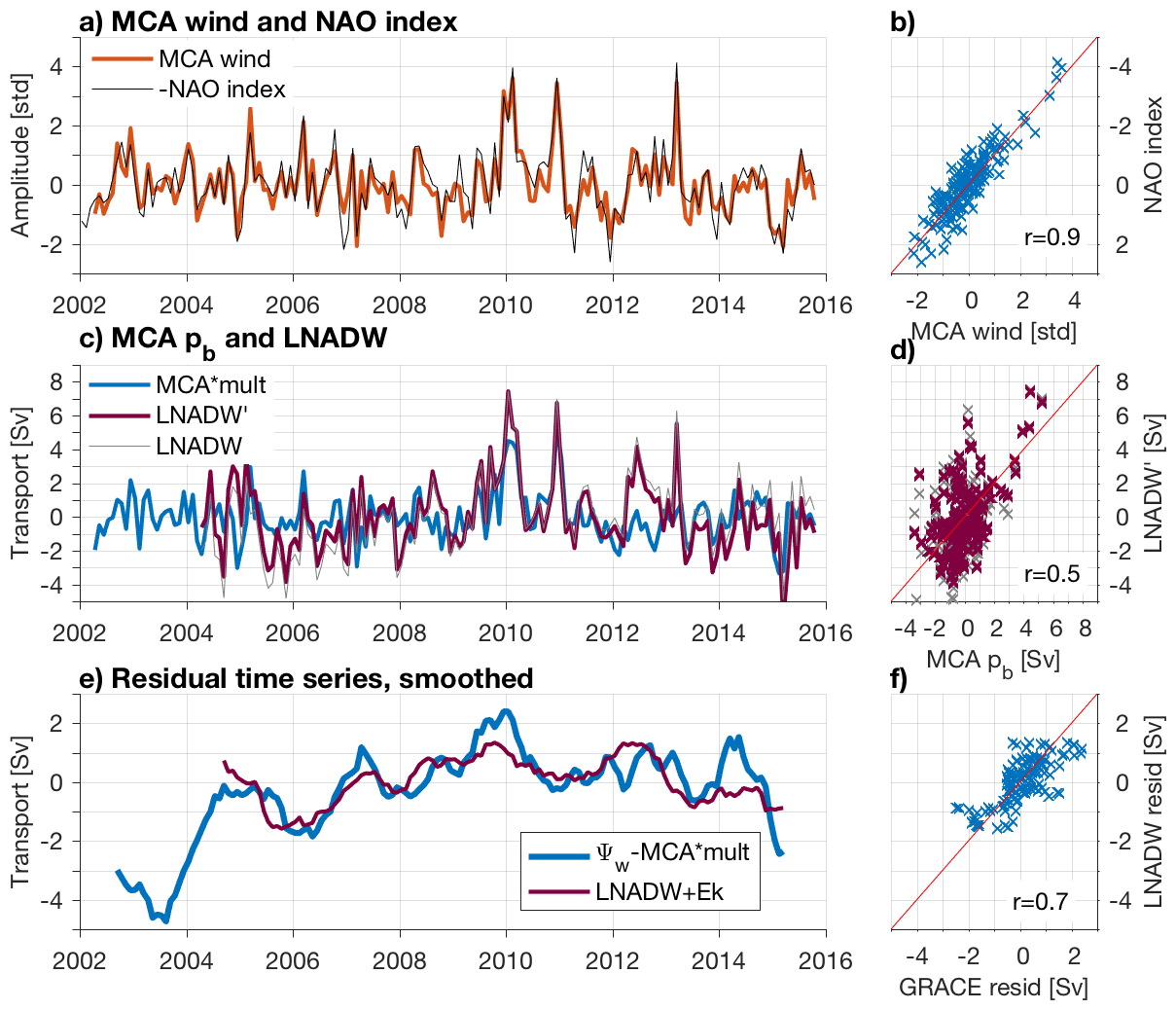}%{tseries_compare_new2.png}
%\centering\includegraphics[width=.7\textwidth]{resid_lowfreq_20180122.png}
\caption{(a) Time series of the wind MCA (red) and monthly NAO index (black) with (b) associated scatter plot and a line with slope -1 (red).   (c) Time series of the MCA amplitude for GRACE $p_b$ in units of equivalent transport and the LNADW transport from RAPID.  The MCA amplitude has been multiplied by 1.23, which is $gH(-P_w)/f$ where $P_w$ is the value at $27^\circ$N, $76.4^\circ$W from the spatial pattern of the MCA (Fig.~\ref{fig:MCA}a).  The LNADW transport is given as original (grey) and in a time series more comparable to the GRACE dataset (LNADW', purple).  For this latter time series, LNADW values were removed where GRACE data were missing (more than 1/3 of the measurement period absent) and then detrended over the RAPID period (April 2004--October 2015). The scatter between the MCA and  LNADW' is shown in (d) with a 1:1 line (red).}
\label{fig:lnadw}
\end{figure}

The time variations in the MCA amplitudes are highly correlated with the monthly NAO index ($r=0.9$, Fig.~\ref{fig:lnadw}a); the bottom pressure amplitudes are less highly correlated with the RAPID transports ($r=0.5$)  but still significant.  Scaling the amplitude time series by $1.4$ Sv/std (determined as $gH_{2k}(-P_w)/f$, where $P_w$ is the value at the 1271 mascons) we find that the magnitude of transports implied from the MCA for ocean bottom pressure is somewhat smaller than those from the \textit{in situ} transports. Removing the wind effect, do residual fluctuations in GRACE capture deep transport variability at RAPID?

\citet{FrajkaWilliams-etal-2016} identified that the surface Ekman transport at $26^\circ$N is not only anti-correlated with the lower NADW transports (3000--5000~m) but also of the same magnitude.  Summing the lower NADW transports and  meridional Ekman transport thus removes the wind influence.  Similarly, subtracting the transport implied by the MCA from the full GRACE data at mascon 1271 gives an estimate of the residual GRACE transports.  Filtering with a 1-year moving average, we find that the residuals are correlated (Fig.~\ref{fig:lnadw}e, f).  While the lower NADW transport at $26^\circ$N is dominated by a monthly-timescale response to large-scale wind stress curl forcing, but that the residual interannual variability is also captured by the GRACE satellite data.

\section{Summary and Discussion}

Release-05 GRACE monthly mascons grids were used to investigate the relationship between the  wind forcing and deep circulation  in the Atlantic.   Using GRACE, we find  that the intra-annual variability in deep transports at $26^\circ$N are part of a basin-scale response to wind stress curl over the intergyre-gyre region ($30^\circ$W, $40^\circ$N).  From \textit{in situ} measurements, the deep ocean response occurs quickly (within 1 day) \citep{FrajkaWilliams-etal-2016}, and primarily in the 3000--5000m layer.  Numerical investigations of  the anomalous transports in the 2009/10 and 2010/11 winters anticipated the deep ocean response found here from observations, which further identify that the ocean response is focused along contours of planetary vorticity ($f/H$).  This paper brings together ocean mass anomalies previously identified using GRACE and wind datasets \citep{Piecuch-Ponte-2014} with the links between GRACE and deep ocean transports at $26^\circ$N \citep{Landerer-etal-2015}. Taken together, we show that the short-timescale reversals in the AMOC at $26^\circ$N are part of a basinscale response to non-local winds.  

%The projection of this barotropic ocean response to wind stress curl onto the baroclinic overturning is then through flow-bathymetry interactions primarily at the western boundary of the Atlantic \citep{Yeager-2015}.

%Using GRACE, we are able to now see that the response at $26^\circ$N is part of a larger, basinscale response.  Not only that, but when removing the wind-forcing effect from the \textit{in situ} and GRACE-derived transports, we find that the interannual variability of residual transport anomalies in the \textit{in situ} lower NADW transports is captured by the GRACE satellite measurements of ocean bottom pressure.

%These findings show the direct response of the deep ocean circulation to wind forcing, and help to place the RAPID AMOC observations at $26^\circ$N into a basin-scale context.  

The reduction of the AMOC strength over the past decade is due in part to longer timescale changes in the lower NADW transports \citep{Smeed-etal-2014}, but these are derived in RAPID as a residual through hypsometric compensation  \citep{McCarthy-etal-2012,FrajkaWilliams-etal-2016}. While these fluctuations have been shown to correlate with bottom pressure gradients on sub-annual timescales \citep{Kanzow-etal-2007,McCarthy-etal-2012}, \textit{in situ} bottom pressure sensors are unable to measure decadal-scale changes due to intrinsic drift \citep{Watts-Kontoyiannis-1990}.  \cite{Landerer-etal-2015} previously used the GRACE estimates of ocean bottom pressure to independently determine the strength of the deep ocean transports from zonal gradients in bottom pressure, but the interannual variations in their time series were dominated by the monthly-timescale wind-fluctuations found here.  We now show that removing the wind effect, the residual interannual variations in GRACE also capture the interannual variations in residual lower NADW transports.  Uncertainty in regional bottom pressure trends from GRACE precludes investigating transport trends further.

%On longer timescales...

This work demonstrates the power of GRACE observations at capturing deep ocean circulation, but must be accompanied by a caveat.  At $27^\circ$N, the western boundary mascons 1271 covaries strongly with the \textit{in situ} bottom pressure sensors (not shown), but this may be due to fortuitous mascons placement at $26^\circ$N or an effect of the steep sidewall of the western boundary bathymetry.  Zonal placement of mascons are adjusted to optimize ocean bottom pressure from GRACE, but each mascon still represents an area of the ocean which is $300\times300$~km.  Within the 1271 mascons, there is substantial variability in bottom pressure records from \textit{in situ} recorders.  Had the variability in this mascons not matched that governing the RAPID transports, the transport relationships found here might have been less favorable.  While the boundary measurements should not affect estimates of large-scale gyre spinup (Fig.~\ref{fig:MCA}a), for transbasin transports, the measurements at the boundary are crucial.

\appendix
\section{Supplementary information}

\acknowledgments
EFW was funded by a Leverhulme Trust Research Fellowship.  The JPL-RL05M GRACE solutions are available via the Physical Oceanography Distributed Active Archive Center (PODAAC) as well as the GRACE Tellus websites (www. grace.jpl.nasa.gov). Data from the RAPID Climate Change (RAPID)/Meridional Overturning Circulation and Heat flux Array (MOCHA) projects are funded by the Natural Environment Research Council (NERC) and National
601 Science Foundation (NSF, OCE1332978), respectively. Data from the RAPID-WATCH MOC monitoring project are funded by the Natural Environment Research Council and are available from \verb+www.rapid.ac.uk/rapidmoc+.

\listofchanges
%%%

\end{document}